\documentclass[12pt]{article}

\pdfoutput=1 
\usepackage{amsmath}
\usepackage{amssymb}
\usepackage{graphicx}

\usepackage{hyperref}
\usepackage{subfig}
\newcount\xrefpos \xrefpos=0
\newcount\yrefpos \yrefpos=0
\newcount\xput \xput=0
\newcount\yput \yput=0

\def\refpos#1 #2 #3{\global\xrefpos=#1 \global\yrefpos=#2
                         \rlap{$\smash{#3}$}}
\def\put #1 #2 #3{\xput=#1 \yput=#2
                  \advance\xput by -\xrefpos
                  \advance\yput by -\yrefpos
                  \rlap{\kern\the\xput truebp
                        \vbox to 0pt{\vss\hbox{$\displaystyle #3$}
                        \kern\the\yput truebp}}}
\def\beginlabels\refpos#1\endlabels{\hbox{$\refpos#1$}}

\newcommand{\be}{\begin{equation}}
\newcommand{\ee}{\end{equation}}
\newcommand{\bea}{\begin{eqnarray}}
\newcommand{\eea}{\end{eqnarray}}
\def\bse{\begin{subequations}}
\def\ese{\end{subequations}}

\def\IZ{\relax\ifmmode\hbox{Z\kern-.4em Z}\else{Z\kern-.4em Z}\fi}

\renewcommand{\d}{\partial}

\renewcommand{\(}{\left(}
\renewcommand{\)}{\right)}

\newcommand{\rootg}{\sqrt{-g}}

\def\bi{\begin{itemize}} \def\ei{\end{itemize}}

\def\({\left(} \def\){\right)}
\def\[{\left[} \def\]{\right]}
\def\be{\begin{equation}}
\def\ee{\end{equation}}
\def\ben{\begin{equation*}}
\def\een{\end{equation*}}
\def\beqa{\begin{eqnarray}}
\def\eeqa{\end{eqnarray}}

\begin{document}

 \begin{center}
 {\Large \bf Chiral Edge Currents in a Holographic Josephson Junction}

\bigskip
\bigskip
\bigskip
\bigskip

\vspace{3mm}

Moshe Rozali\footnote{email: rozali@phas.ubc.ca} and Alexandre Vincart-Emard\footnote{email: ave@phas.ubc.ca} 

\bigskip\medskip
\centerline{\it Department of Physics and Astronomy}
\smallskip\centerline{\it University of British Columbia}
\smallskip\centerline{\it Vancouver, BC V6T 1Z1, Canada}

 \bigskip\bigskip\bigskip

 \end{center}

\abstract{We discuss the Josephson effect and the appearance of dissipationless edge currents in a holographic Josephson junction configuration involving a chiral, time-reversal breaking, superconductor in 2+1 dimensions. Such a superconductor is expected to be topological, thereby supporting topologically protected gapless Majorana-Weyl edge modes. Such modes can manifest themselves in chiral dissipationless edge currents, which we exhibit and investigate in the context of our construction. The physics of the Josephson current itself, though expected to be unconventional in some non-equilibrium settings, is shown to be conventional in our setup which takes place in thermal equilibrium.  We comment on various ways in which the expected Majorana nature of the edge excitations, and relatedly the unconventional nature of  topological Josephson junctions, can be verified in the holographic context.}

\newpage

\section{Introduction and Summary}

The physics of topological insulators and superconductors has become a central topic in modern condensed matter physics (for reviews see \cite{qi,hassan}). Many of the interesting phenomena exhibited in such materials follow from the existence of topologically protected gapless edge modes. For topological superconductors, these are expected to be chiral Majorana modes. The search for such Majorana excitations in various condensed matter systems is currently an intense experimental effort (for a review see \cite{majorana}).

Topological superconductivity is expected to arise in time-reversal breaking superconductors, with a ``p+ip" order parameter symmetry, which we refer to here as chiral superconductors. Experimentally, such topological superconductivity might arise intrinsically, for example in the Strontium Ruthenate $Sr_2RuO_4$ (see e.g. \cite{pwave1,pwave2} for reviews), or by proximity effect (following the suggestion of Fu and Kane \cite{proximity}). That system was analyzed by Green and Read \cite{Read:1999fn}, who demonstrated the existence of Majorana-Weyl fermions propagating on the edges of a two-dimensional chiral superconductor\footnote{See however \cite{moller1} for a recent null experimental result in $Sr_2RuO_4$.}. A particularly clear construction of the edge modes as Andreev bound states can be found in \cite{stone}.

In this note we use the tools of gauge-gravity duality to investigate the topological nature of the holographic superconductor. As we will see, the  manifestation of topology comes in the form of chiral edge excitations which manifest themselves as  chiral currents localized at edges of the superconductor. To this end, we construct a gravity solution that exhibits the basic phenomena associated with topological superconductivity, namely topologically protected edge modes and spontaneously generated edge currents.  Indeed, after the observation of \cite{Gubser:breaking} that black holes can be unstable to scalar condensation, an s-wave superconductor was constructed  in \cite{Hartnoll:2008}. Holographic duals to p-wave superconductors were constructed in \cite{Pufu:2008}, and the model we are using here, an holographic dual to a chiral superconductor, was constructed in \cite{Gubser:2008}. We review that construction in section 2 below\footnote{The model we discuss here supports competing orders, indeed the p-wave order parameter \cite{Pufu:2008} is thermodynamically preferred in this model. Our Josephson junction is therefore an idealized configuration, but is nevertheless an interesting probe of the time-reversal breaking holographic superconductor. We expect that the features we uncover  here, to do with topological structures, are insensitive to the phase structure of the full model.}.

Since much of the new interesting physics associated with topological superconductivity has to do with edges and interfaces, we construct a Josephson junction involving the holographic chiral superconductor.
Holographic Josephson junctions were constructed first in \cite{Horowitz:2011ch} (see also \cite{Kiritsis:2011zq,Wang:2011ri,Wang:2012yj} for other configurations). Our work will focus on building an S-N-S holographic Josephson junction for the holographic chiral superconductor (S) for which the weak link is a normal metal (N). The construction of the gravity solution involves the numerical solution of a set of partial differential equations, details of which are presented in section 2.

One of the dramatic manifestations of the topologically protected gapless modes are  spontaneously generated dissipationless currents, localized at the edges of a topological superconductor. The relation between the edge currents, edge states and gauge invariance is explained in \cite{stone}. Since Josephson junctions involve two such interfaces between topological and non-topological materials, we expect to find counter-propagating currents, one on each interface. Such currents are clearly visible in our setting  and we discuss their features in section 3 below.  We find that up to small corrections, the strength of the edge currents is determined by the jump of the order parameter amplitude across the interface between the superconducting material and the weak link.

The counter-propagating edge currents we observe are independent of each other (for wide enough junctions), and would exist for a single isolated interface as well. They are indicative of chiral gapless edge modes localized on such interfaces\footnote{For the existence of a charge current, at least two Majorana-Weyl fermions are required. We thank Andreas Karch for a related discussion.}. The full Josephson junction has a pair of these modes, a feature which is speculated to be responsible for some unusual properties of the topological Josephson junction. Therefore, in section 4 we turn to examine the Josephson current in our junction.

Anomalies in the current-phase relation in such ``unconventional" Josephson junctions were reported in \cite{expt1}, but a more recent direct measurement reveals a conventional relation \cite{expt2}. While the physics of such junction is expected to be unconventional, in that it is $4 \pi$ periodic in the phase across the weak link \cite{kitaev}, equilibrium configurations might still exhibit the conventional $2 \pi$ periodicity. Other attempts to discover unconventional periodicity as a signature of the aforementioned pair of Majorana bound states include the AC Josephson effect \cite{ac},  Josephson junctions in magnetic fields \cite{magneto}, current noise measurements \cite{noise} or unconventional Shapiro steps \cite{shapiro}.

In section 4 we exhibit the details of the Josephson effect in our holographic construction. We find conventional results, fairly similar to the s-wave case reported in \cite{Horowitz:2011ch}. The current-phase relation is $2 \pi$ periodic and the maximal current decays exponentially as the width of the junction increases (as opposed to the power law decay observed in \cite{expt1}). Furthermore, the temperature dependence of the critical current and the coherence length are also found to be fairly conventional. We conclude that our setup, which takes place in thermal equilibrium, is  thus insensitive to the unconventional features expected to arise from the presence of gapless Majorana modes.

We conclude in section 5 with outlook and directions for future work. In particular, we outline some calculations that would verify the existence of gapless Majorana modes and exhibit the expected doubled periodicity of the physics in the Josephson junction. We hope to report on such results in the near future.

\bigskip
\section{Setup and Solutions}
\label{sec_setup}

Our discussion of the time-reversal breaking holographic superconductor \cite{Gubser:2008} follows the conventions of \cite{Roberts:2008}. Let us consider the following action:
\be \label{action} S = \int d^4 x \rootg \[ R + \frac{6}{L^2}- \frac{1}{4g^2} (F_{\mu \nu}^a)^2 \] \ee
where $F_{\mu \nu}^a = \d_\mu A_\nu^a - \d_\nu A_\mu^a + \epsilon^{a b c} A_\mu^b A_\nu^c $ is the field strength tensor for an $SU(2)$ gauge field, and $\epsilon^{a b c}$ is the totally antisymmetric tensor, with $\epsilon^{123} = 1$.  The gauge field can be conveniently expressed as a matrix-valued one form: $A = A_\mu^a \tau^a dx^\mu$, where $\tau^a = \sigma^a/2i$,  $\sigma^a$ being the usual Pauli matrices. It follows that $[ \tau^a, \tau^b] = \epsilon^{a b c} \tau^c$. 

We will be working in the probe approximation, thereby neglecting the backreaction of the gauge field on the metric. In the current model, the probe approximation, controlled by the ratio of the Newton's constant to the gauge coupling, breaks down at sufficiently low temperatures. However, though adding backreaction should be straightforward, this is unnecessary for the effects we  are interested in, and we will restrict ourselves to working in a fixed gravitational background.

Specifically, we choose the metric to be the asymptotically $AdS_4$  planar Schwarzschild black hole:
\begin{align} ds^2 = - h(r) dt^2 + \frac{dr^2}{h(r)} + r^2 (dx^2 + dy^2) \nonumber \\
 \mathrm{where}Ã\;\;\;\;\; h(r) = \frac{r^2}{L^2}\(1-\frac{r_0^3}{r^3}\) \end{align}
 and $L$ and $r_0$ are the AdS and horizon radii, respectively. Such a black hole has Hawking temperature 
\be T = \frac{1}{4 \pi} \frac{dh}{dr}\biggr\rvert_{r=r_0}= \frac{3 r_0}{4 \pi L^2} \ee
Scaling symmetries further enable us to work in units in which $L=1$ and set $r_0=1$. This corresponds to measuring all dimensional quantities in units of temperature.

To understand the symmetry structure of our ansatz, it is useful to define complex coordinates 
\be \zeta = \frac{x+iy}{\sqrt{2}} \;\;\; \mathrm{and} \;\;\; \tau^{\pm} = \frac{\tau^1 \pm i \tau^2}{\sqrt{2}} \ee
The ansatz for the spatially homogeneous $p+ip$ superconductor is given by \cite{Gubser:2008,Roberts:2008}
\be A = \Phi \; \tau^3 dt + w \tau^- d\zeta + w^* \tau^+ d\bar{\zeta}  \ee
Here $\Phi$ breaks the $SU(2)$ symmetry  explicitly to an Abelian subgroup at the short distance scale of the chemical potential $\mu$, and $w$ is the order parameter which breaks the $U(1)$ symmetry spontaneously  at a much longer distance scale. Note that in these conventions a $U(1)$ gauge transformation is a phase rotation of the complex order parameter $w$. In the homogeneous case $w$ can be chosen to be everywhere real, but with inhomogeneities  this is no longer the case. In particular the phase difference across the Josephson junction is an interesting gauge invariant quantity.

The order parameter $w$ is invariant under a combination of spatial rotations and gauge transformations, thus the superconductor is isotropic. Since $w$ is complex, time-reversal is broken spontaneously. As we see below, this has interesting consequences for the physics probed by the Josephson junction in this system.

In order to build a holographic Josephson junction, the fields must have spatial dependence. We model a Josephson junction by choosing an appropriate profile for the chemical potential $\mu(x)$, as described below. The fields then all depend on the spatial coordinate $x$ and the radial coordinate $r$. Our ansatz for a $p+ip$ Josephson junction is then
\begin{align} A &= \Phi \; \tau^3 dt + w \tau^- d\zeta + w^* \tau^+ d\bar{\zeta}\nonumber \\
&+ A_x \tau^3 dx + M_y \tau^3 dy + A_r \tau^3 dr \label{ansatz} \end{align}

We are using a somewhat mixed notation for the spatial directions, where the chiral nature of the order parameter is most clearly exhibited using the $\zeta$ coordinate defined in (4). Note the presence of the field $M_y$ which, unlike other instances of holographic Josephson junctions, cannot be eliminated using symmetries. This field will encode the presence of dissipationless edge currents, which we discuss below.

Following \cite{Horowitz:2011ch} we choose to work in terms of gauge invariant combinations. If $w=|w| e^{i \theta}$, those are $w\equiv |w|$, $M_y$ and $M_\mu=A_\mu -\partial_\mu \theta$ for $\mu={x,r}$. Our ansatz yields the following system of 5 coupled non-linear elliptic PDEs, 
\begin{align}
\partial_r^2 \Phi + \frac{2}{r} \partial_r \Phi + \frac{\partial_x^2 \Phi}{r^2h}  - \frac{2 w^2 \Phi}{r^2 h} &=0 \nonumber \\
\partial_r^2 w + \frac{h^\prime}{h} \partial_r w + \frac{\partial_x^2 w}{2r^2h}  - \frac{3 w \partial_x M_y}{2 r^2 h}  + \frac{w \Phi^2}{h^2} - \frac{w^3}{r^2 h} - \frac{(M_x^2+M_y^2)w}{2r^2h} - M_r^2 w  &= 0  \nonumber \\
\partial_r^2 M_x + \frac{h^\prime}{h}(\partial_r M_x - \partial_x M_r) - \partial_r \partial_x M_r - \frac{M_x w^2}{r^2h}  &=0   \nonumber \\
\partial_r^2 M_y + \frac{h^\prime}{h}\partial_r M_x +\frac{\partial_x^2 M_y}{r^2 h} + \frac{3 \partial_x(w^2)}{2 r^2 h} - \frac{M_y w^2}{r^2h}  &=0  \nonumber\\
\partial_x^2 M_r - \partial_r \partial_x M_x - 2 M_r w^2 &=0
\end{align}
and an additional constraint:
\be  \partial_r \left(h M_r w^2 \right) + \frac{1}{2 r^2} \partial_x \left( w^2  M_x \right) = 0 \ee
We thus need to choose boundary conditions such that the constraint is satisfied at the boundaries of the integration domain.

Next we discuss the boundary conditions satisfied by our fields. The boundary conditions at the horizon are determined by requiring regularity and satisfying the constraint. That is, when expanding the equations of motion and constraint near the horizon, divergent terms arise which we require to cancel. At the spatial boundaries we impose Neumann boundary conditions on all fields. Near conformal infinity the fields behave asymptotically as
\begin{align}
\Phi(r,x) &= \mu(x) - \frac{\rho(x)}{r} + \cdots \nonumber\\
w(r,x) &= w^{(1)}(x) + \frac{ w^{(2)}(x) }{r} + \nonumber \cdots \\
M_x(r,x) &= v_x(x) + \frac{J_x}{r}+ \nonumber \cdots \\
M_y(r,x) &= v_y(x) + \frac{J_y(x)}{r}+\nonumber  \cdots \\
M_r(r,x) &=  O\left( \frac{1}{r^3} \nonumber \right) 
\end{align}

We will input $\mu(x)$ to model a Josephson junction, and choose the condensate $w$ to be normalizable, $w^{(1)}(x)=0$. The current in the x-direction $J_x$ is constant by the continuity equation and we choose it to be one of the parameters of our solutions. The conjugate quantity $v_x(x)$ is then read off the solution and encodes the phase difference across the junction. Finally, we set $v_y(x)=0$ as there is no applied voltage in the transverse direction $y$, and read off the spontaneously generated  transverse current $J_y(x)$ from the solution.

To model a Josephson junction we need to choose the profile $\mu(x)$ appropriately. In the case of homogeneous superconductors, the scale invariant quantity to consider is $T/\mu$, i.e. changing the temperature is equivalent to  changing $\mu$. This is no longer the case in the spatially inhomogeneous case: while our chemical potential is spatially varying, the  temperature is still constant. Instead, to measure the temperature in a scale invariant way we use the scale invariant quantity $T/T_c$, where the critical temperature $T_c$ is proportional to $\mu_\infty \equiv \mu(\infty)$. Our simulations set the proportionality relation to be
\be T_c \approx 0.065 \; \mu_\infty \ee
Since we now change the temperature by varying $\mu$, there is a corresponding critical chemical potential $\mu_c$ below which the condensate vanishes. We thus need a profile with the following features:

\begin{equation*}
  \begin{cases}
    \mu(x)<\mu_c & \text{in the normal metal phase,} \; -\frac{\ell}{2} < x < \frac{\ell}{2} ;\\
    \mu(x)>\mu_c & \text{otherwise for the superconducting phases}.
  \end{cases}
\end{equation*}
As in \cite{Horowitz:2011ch}, a profile that satisfies these conditions is 

\be \mu(x) = 
 \mu_\infty \left[1-Ã\frac{1-\epsilon}{2 \tanh(\frac{\ell}{2 \sigma})} \left\{ \tanh\left( \frac{x+\frac{\ell}{2}}{\sigma} \right) - \tanh\left( \frac{x-\frac{\ell}{2}}{\sigma} \right) \right\} \right] \ee
where $\mu_\infty>\mu_c$ is the maximal height of the chemical potential. 
The parameter $\sigma$ controls the steepness of the profile, whereas $\epsilon$ controls its depth -- the chemical potential inside of the normal phase is typically $\mu_0 \equiv \epsilon \mu_\infty$. Moreover, it is convenient to work with compactified variables  $z=1/r$ and $\bar{x}=\tanh(\frac{x}{4 \sigma})/\tanh(\frac{p}{4 \sigma})$, where $p$ is the length of the $x$-direction.

Pseuso-spectral collocation methods on a Chebyshev grid were used to discretize the above equations. The resulting equations were solved using the Newton iterative method. One key characteristic of the pseudo-spectral methods is their exponential convergence in the size of the grid used, which we have confirmed explicitly for our solutions. The solutions used in this paper  were produced using a grid of 41 points in both the radial and spatial direction, yielding  an estimated maximal error of about $10^{-4}$ in the local value of all functions. 

\begin{figure}[b!] 
\centering
    \includegraphics[width=16cm]{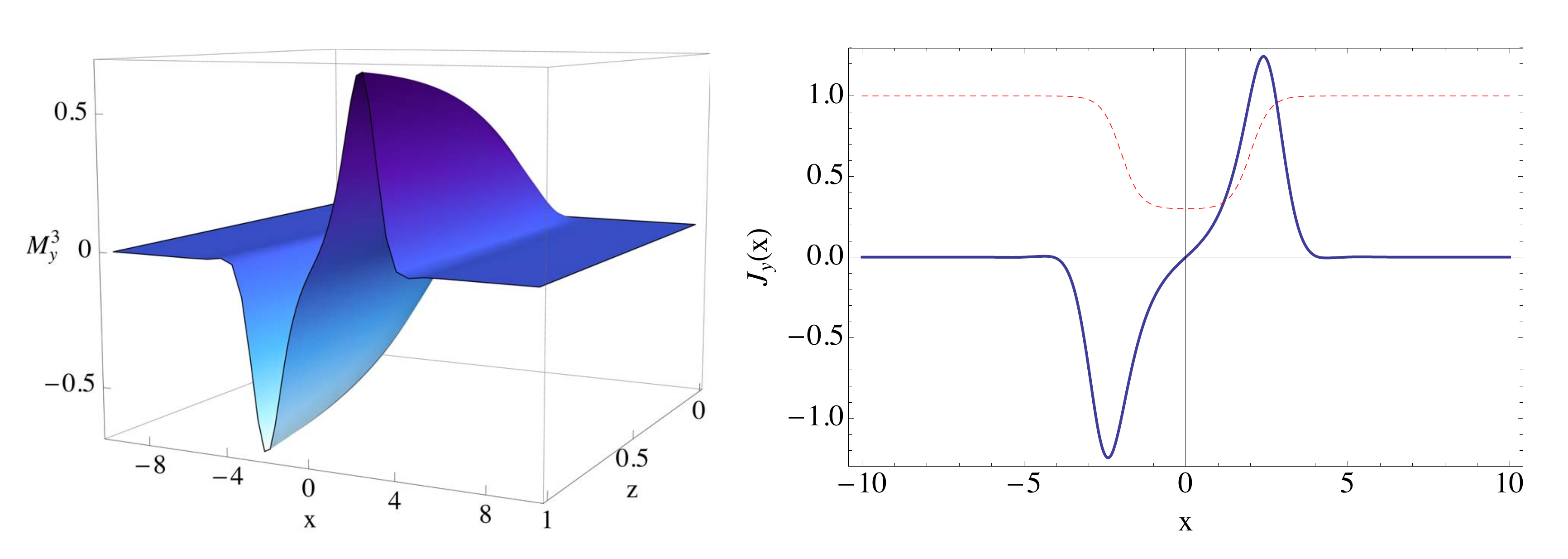}
    \caption[]{The left plot displays the solution for the gauge field $M_y^3$. The right plot shows the resulting boundary current $J_y(x)$ (in blue), as well as the rescaled chemical potential profile $\mu(x)/\mu_\infty$ (in red). The solution corresponds to $\mu_\infty =7$, $\ell=4$, $\sigma=0.6$ and $\epsilon=0.3$. }
\label{fig_Jedge_Ay}
\end{figure}
\bigskip
\section{Chiral Edge Currents}
\label{sec_edge}

We start by discussing the phenomenon unique to the time-reversal breaking chiral superconductor, the existence of edge currents. The existence of dissipationless chiral edge currents is indicative of gapless chiral modes living on an interface between the superconductor and the normal state. In a Josephson junction configuration there are two such interfaces and therefore two independent counter-propagating modes. 
In this section we focus on aspects of these modes that are localized at each interface separately, which would exist in a simpler domain wall geometry. In the next section we turn to discuss aspects of the physics more specific to a Josephson junction configuration with two such interfaces.

To be more concrete, the introduction of a gauge field $M_y^3$ makes it possible to measure a current $J_y(x)$ propagating in the y-direction. We have specified $M_y^3(r=\infty,x)=0$ so that the system has no applied voltage that would drive a current in the y-direction, thus this is a dissipationless current flowing without resistance. Under such conditions, the field $M_y^3$ would vanish everywhere for a $p$-wave order parameter, but has a non-trivial profile in the $p+ip$ case. As shown in figure \ref{fig_Jedge_Ay}, this current is localized at the interfaces of the superconducting and normal phases, travelling in opposite directions with equal magnitude. We have checked that the strength of the current is independent of the Josephson phase (or equivalently, the strength of the Josephson current) and the width of the junction (for sufficiently wide junctions). Therefore the currents on both interfaces are local effects independent of each other.

\begin{figure}[t!] 
\centering
    \includegraphics[scale=0.3]{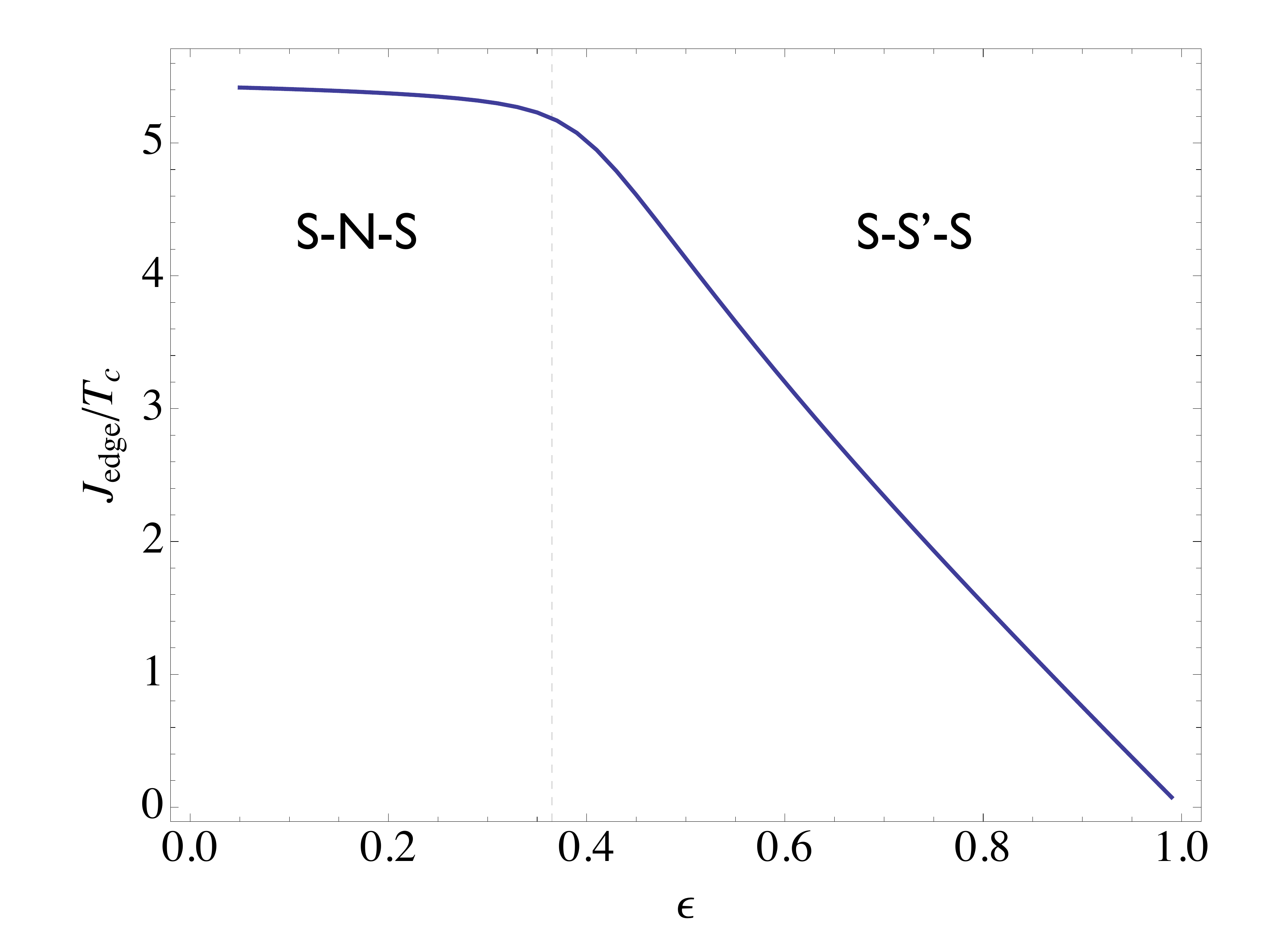}
    \caption[]{The data above corresponds to $\mu_\infty=10$. As we increase the depth $\epsilon$, we observe a linear decrease in the edge current value happening around the dashed line at $\epsilon_c = \mu_c/\mu$, which is the critical depth at which the weak link becomes superconducting. At $\epsilon=1$, $J_{\text{edge}}$ goes to 0 since $\mu(x)$ becomes homogeneous.}
\label{fig_edge_depth}
\end{figure}

\begin{figure}[t!] 
\vspace{-2cm}
\centering
\includegraphics[scale=0.4]{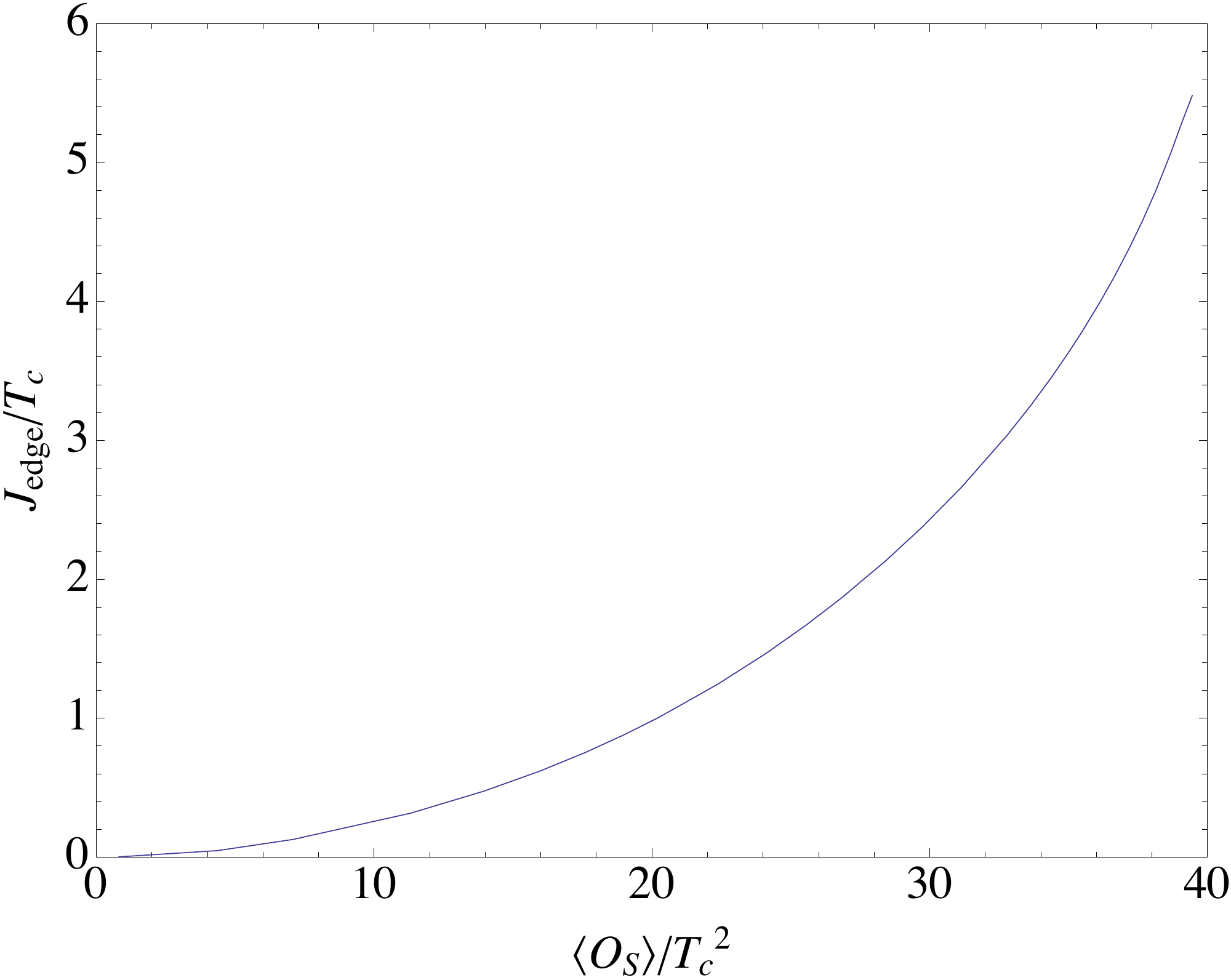}
   \caption[]{This plot illustrates the dependence of the edge current on the amplitude of the order parameter in the superconducting phase for $\ell=4$, $\sigma=0.6$ and $\epsilon=0.3$. The curve fits $J_{\text{edge}} \sim \langle \mathcal{O_S} \rangle ^\alpha$ with $\alpha \simeq 2$.}
\label{fig_edgeOS} 
\end{figure}

The quantity of interest is the total current per unit area, defined as follows:
\be J_{\text{edge}} \equiv \int_0^{\infty} J_y(x) dx \ee

We focus on this quantity as it is independent of details of the interface profile such as the steepness, parametrized by $\sigma$ above. Furthermore, we find that the edge current is essentially constant when the weak link is a normal metal, independent of the relative depth parameter $\epsilon$. However, when the weak link becomes superconducting, the current decreases as we increase $\epsilon$, eventually vanishing when the solution is perfectly homogeneous.

This dependence on the relative depth is shown in Figure \ref{fig_edge_depth}. It indicates that the current is controlled by the jump in the amplitude of the order parameter across the interface between the superconducting material and the weak link. Indeed, in figure \ref{fig_edgeOS} we plot the dependence on the edge current on the order parameter in the superconducting phase $\langle \mathcal{O}_S\rangle$  (choosing $\epsilon$ such that the weak link is always at the normal state, i.e. has approximately zero condensate). The edge current depends on the magnitude of the order parameter through a power law relationship $J_{\text{edge}} \sim \langle \mathcal{O}_S \rangle ^\alpha$ with $\alpha$ ranging between 2.04 and 2.12 for different choices of parameters.

\begin{figure}[b!] 
\centering
\includegraphics[scale=0.4]{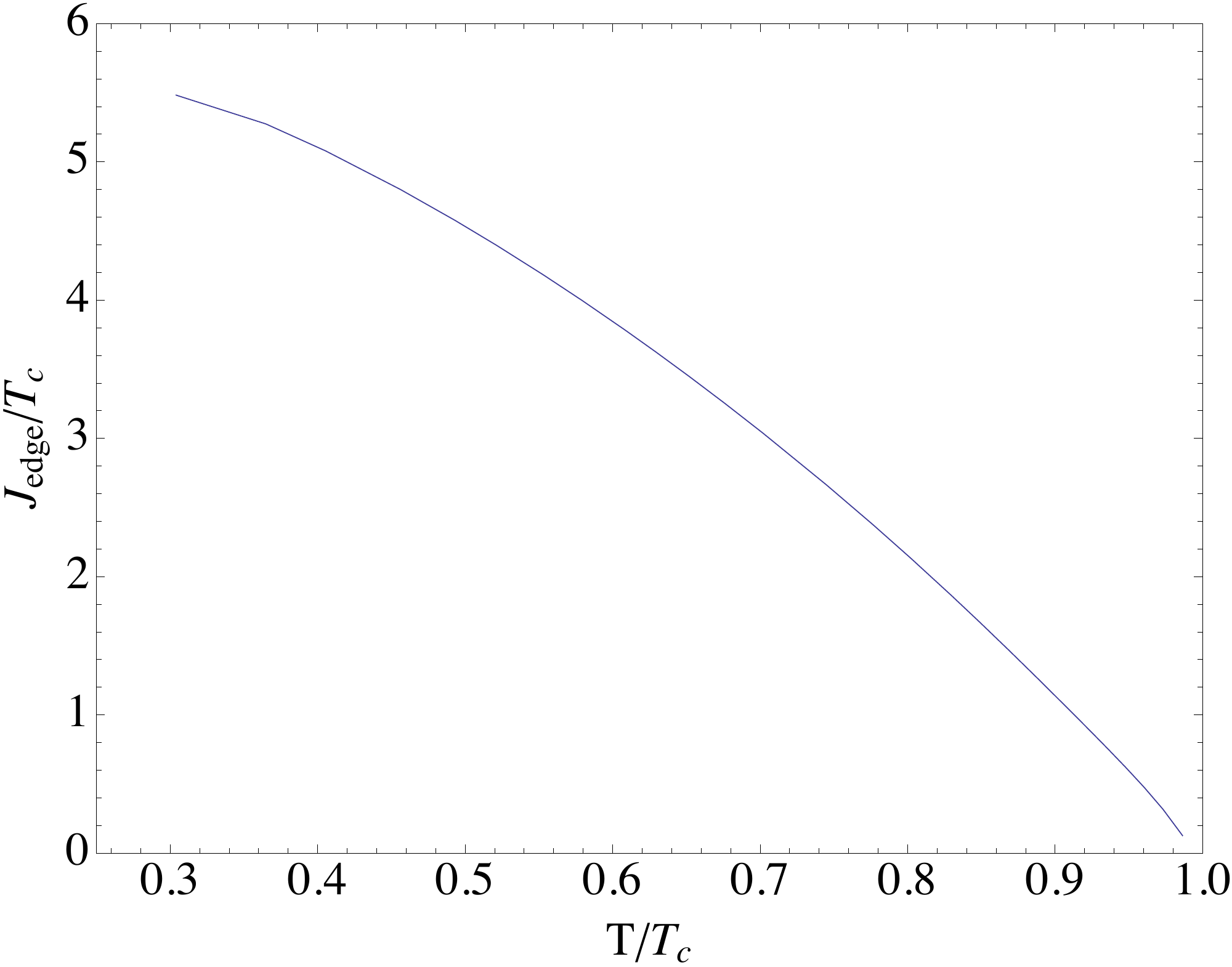}
   \caption[]{The temperature dependence of the edge current is pictured for $\ell=4$, $\sigma=0.6$ and $\epsilon=0.3$.}
\label{fig_edgetemp} 
\end{figure}

It is also natural to examine the temperature dependence of the edge current, which we plot in  figure \ref{fig_edgetemp}. We see that the dominant change in the edge current as we change the temperature comes through the change in the amplitude of the order parameter. As expected the edge current vanishes at the critical temperature $T_c$.

\bigskip
\section{Josephson Currents}
\label{sec_josephson}

The Josephson effect is a macroscopic quantum phenomenon in which a  dissipationless current flows across a weak link between two superconducting electrodes, in the absence of an external applied voltage. Rather, it is the gauge invariant phase difference across the junction that is responsible for the current. Following \cite{Horowitz:2011ch}, we will consider S-N-S Josephson junctions, for which the weak link is a non-superconducting (``normal") metal, as described above. We also make some comments on the S-S'-S case, in which the weak link is superconducting. For a discussion of an  S-I-S weak link, see \cite{Wang:2012yj}.

\begin{figure}[b!] 
\center
    \includegraphics[scale=0.5]{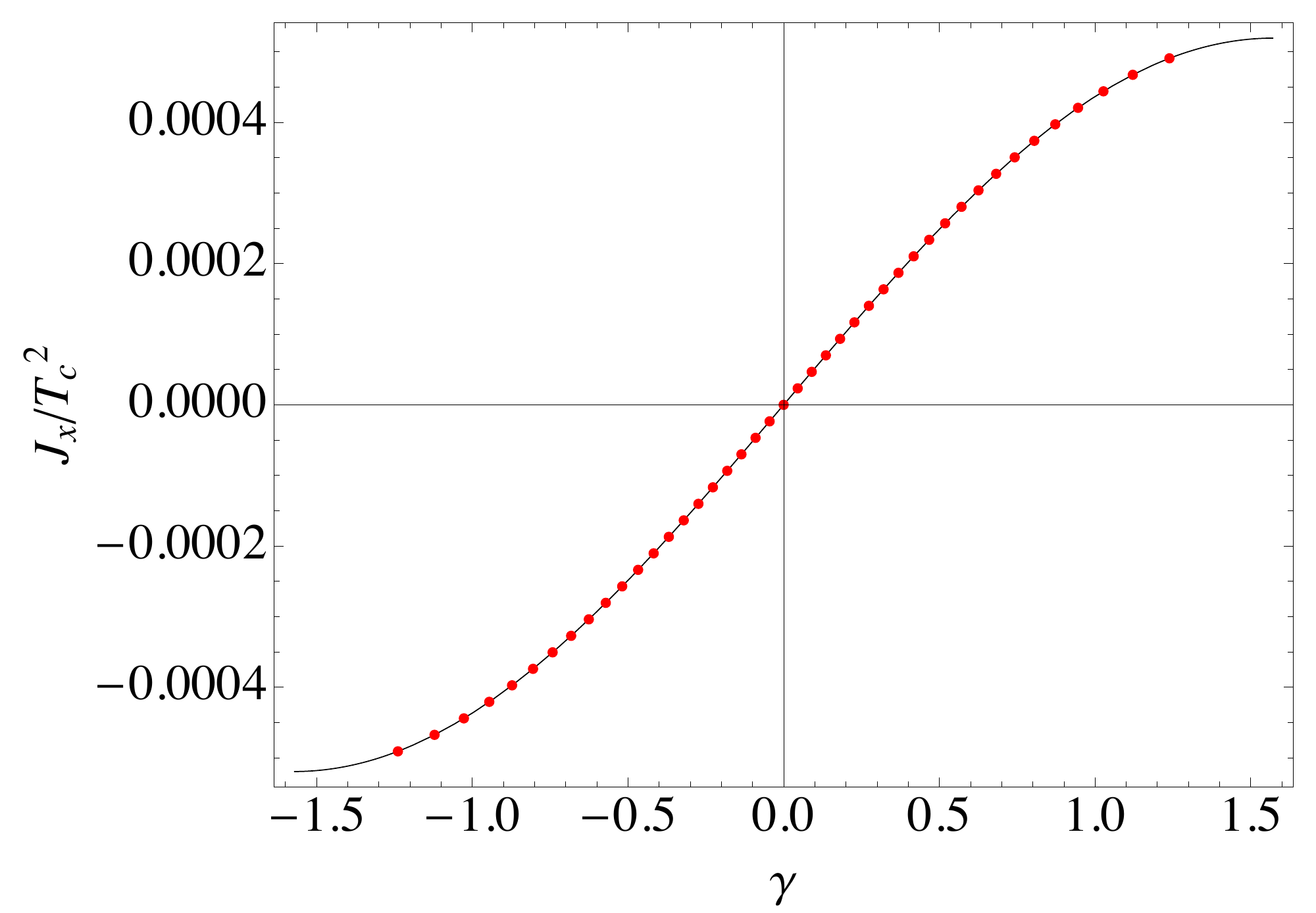}
    \caption[]{This graph, produced with $\mu_\infty = 10$, $\ell = 4$, $\sigma = 0.4$ and $\epsilon = 0.05$, shows the agreement of our data with the expectation (\ref{phase_eq}). The solid line, describing the best fit curve to our data (in red), is the curve  $J_x = 5.2  \times 10^{-4} \sin(0.9986 \gamma)$.}
\label{fig_josephson}
\end{figure}

The Josephson current flowing across the junction has the expected form
\be J_x = J_{\mathrm{max}} \sin \gamma \label{phase_eq} \ee
where the gauge invariant phase difference across the junction is obtained from the solution as
\be \gamma = - \int_{-\infty}^{\infty} dx \left[v_x(x)-v_x(\pm \infty)\right]  \label{jcurrent} \ee

\begin{figure}[h!] 
\centering
    \includegraphics[scale=0.35]{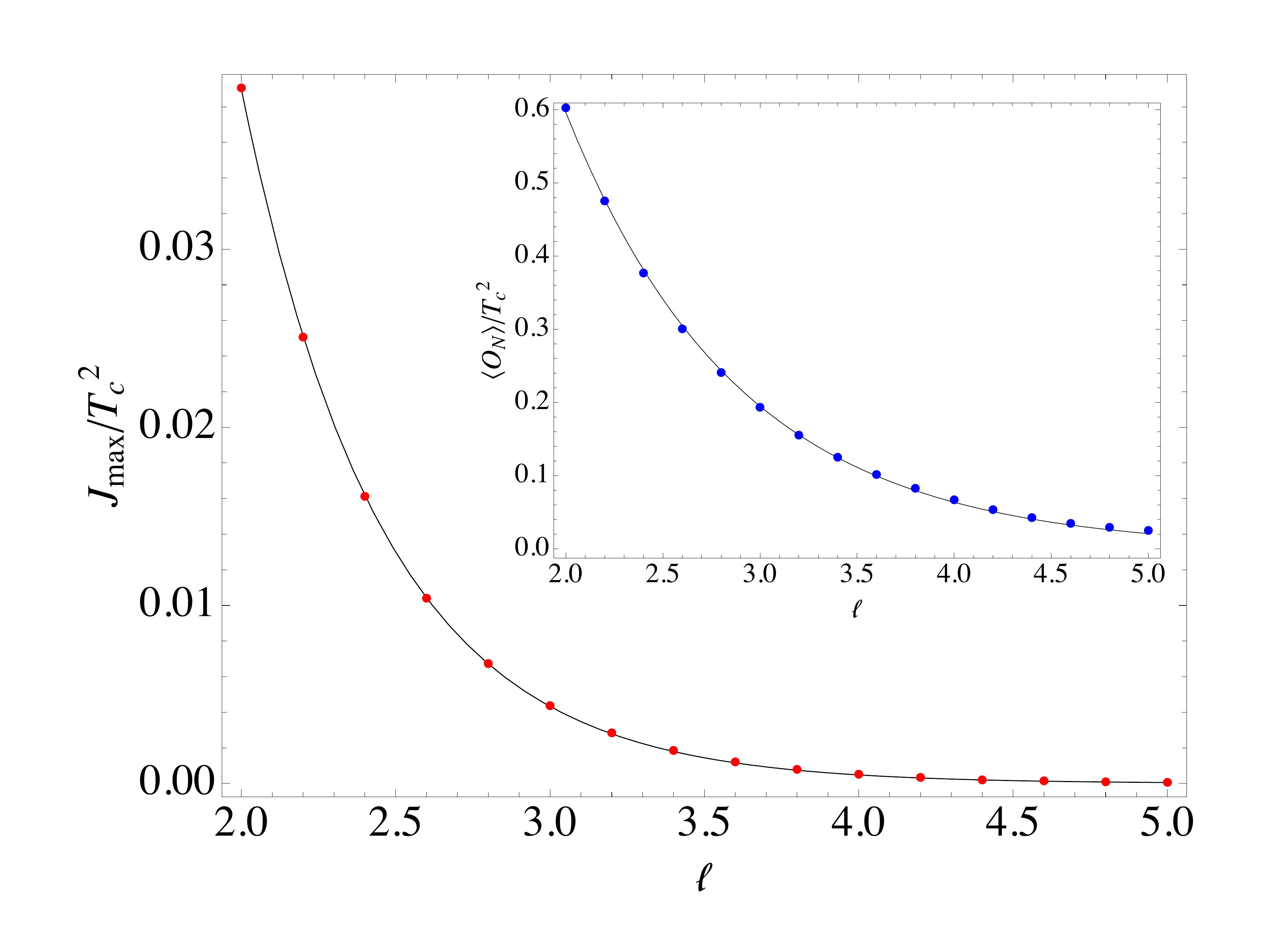}
    \caption[]{The relationship of the critical current and of the order parameter as the weak link length grows larger is illustrated in these two plots. Both sets of data are fitted to a decaying exponential and independently yield the same coherence length, up to a difference of 1.8\%. We used $\mu_\infty=10$, $\sigma=0.4$ and $\epsilon=0.05$ for both plots.}
\label{fig_joswidth}
\end{figure}

Figure \ref{fig_josephson} has been obtained by computing $\gamma$ for multiple solutions corresponding to different inputs $J_x$; it clearly demonstrates the expected dependence of the Josephson current on the phase difference.

Another interesting feature of the critical current $ J_{\mathrm{max}}$ is that it decays exponentially when the width $\ell$ of the weak link increases, i.e. it obeys a  relation of the form\footnote{Note that we have switched from the numerically convenient conventions of measuring dimensional quantities in units of the (varying) temperature $T$, to the more physical conventions of measuring those quantities in terms of the {\it fixed} temperature $T_c$.} 
\be \frac{J_{\mathrm{max}}}{T_c^2} = A_J e^{- \frac{\ell}{\xi}} \ee 
for $\xi \ll \ell$. Additionally, the order parameter at the center of the junction has a similar behaviour:
\be \frac{\langle \mathcal{O}_N \rangle}{T_c^2} = A_{\mathcal{O}} e^{-\frac{\ell}{2 \xi}} \ee
where $\langle \mathcal{O}_N \rangle$ is the magnitude of the order parameter in the normal phase ($x=0$), and the junction has no current. Both of these relations are due to the proximity effect, the leakage of the superconducting order into the normal state. Note then that the coherence length $\xi$ should be the same in both cases. The results of our numerics show remarkable precision, yielding $\xi \approx 0.4547$ for the Josephson current, and $\xi \approx 0.4468$ for the order parameter: a difference of only 1.8\%. See for instance Figure \ref{fig_joswidth}.

\begin{figure}[t!] 
\centering
    \includegraphics[scale=0.3]{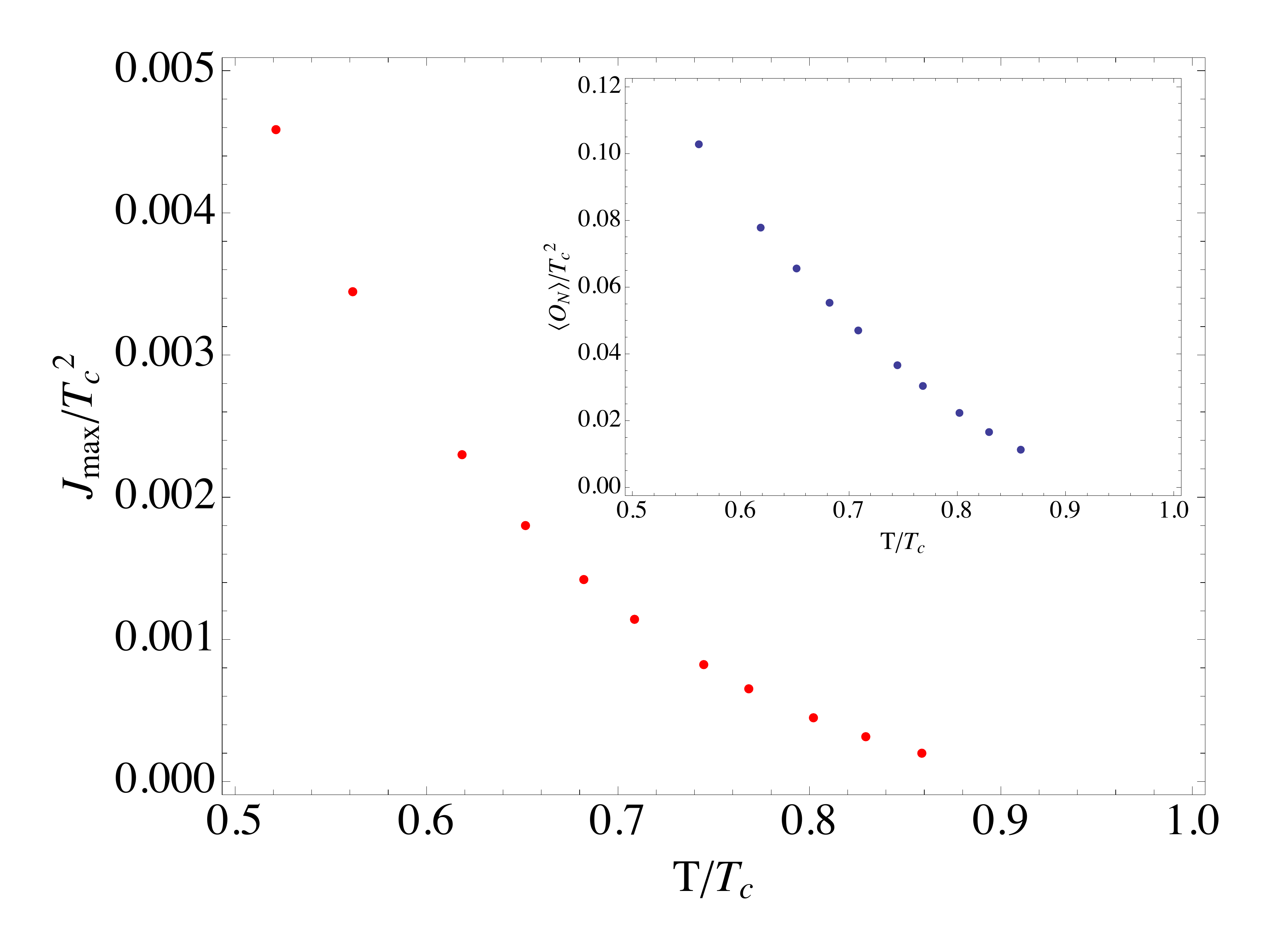}
    \caption[]{The dependence of the coherence length $\xi$ , expressed through the  critical current or order parameter $\langle \mathcal{O}_N\rangle$ (in the inset), on temperature. Near the critical point, the critical current follows a power law characteristic of the S-N-S junction, but with fairly large corrections. The parameters for the chemical potential used to produce these figures are $\ell = 4$, $\sigma=0.6$ and $\epsilon=0.3$.}
\label{fig_jostemp}
\end{figure}

The coherence length has an interesting temperature dependence, plotted in figure \ref{fig_jostemp}. While the plot does not have a simple fit, it has the expected behaviour at $T\rightarrow T_c$, where it vanishes due to the disappearance of superconductivity. Near the critical temperature the critical current is expected to vanish as \cite{Barone}:
\be J_{\text{max}}(T) \propto ( T_c-T)^\beta \;\;\; \text{near} \;\;\; T_c\ee
For a conventional s-wave superconductor, and junctions wide compared to the coherence length, a quadratic dependence ($\beta=2$) is characteristic of the S-N-S junction, whereas different critical exponent are expected for other types of weak links (for example for the S-I-S junction $\beta=1$). Interestingly, our results presented in figure \ref{fig_jostemp} indicate that $\beta \sim 2.52$; furthermore the  exponent $\beta$ also depends on the steepness and depth of the chemical potential profile. While $\beta$ is closest to the critical exponent of the S-N-S junction, the corrections are fairly large. Those corrections are probably related to our setup having a varying chemical potential. It would be interesting to reproduce the expected scaling with a more conventional setup of a Josephson junction for which the chemical potential is spatially homogeneous.

\bigskip
\section{Conclusions}
In this paper we have started the investigation of Majorana bound states in the holographic context. We have discussed the dissipationless edge currents which are an indirect evidence for such modes. Additionally, we have constructed a Josephson junction involving a topological chiral superconductor, and probed the physics of the Josephson effect. The results we obtained are consistent with a conventional effect, with $2 \pi$ periodic current-phase relation, and exponential decay of the current with the junction width. 

These results support  the expectation that though the physics is $4 \pi$ periodic, an unconventional periodicity will not be visible in thermal equilibrium. The presence of Majorana modes corresponds to having two states which are exchanged upon a $2 \pi$ phase rotation. However, in equilibrium the Josephson current receive contribution from both states, weighted according to their Boltzmann weight. This, thermal equilibrium results are expected to exhibits conventional periodicity, consistent with what we find,

It would be interesting to continue this investigation with the goal of displaying more direct signatures of the Majorana bound states. One such direct signature would be in the Andreev scattering off a superconducting interface -- bound states can be then seen in analyzing the phase shift. Furthermore, one can construct holographically a non-equilibrium configuration which is expected to exhibit the unconventional periodicity associated with Majorana bound states.  We hope to report on the result of such investigations in the near future.

\bigskip
\section*{Acknowledgements}

We thank Marcel Franz, Benson Way and Fei Zhou for useful conversations. M.R thanks Centro de Ciencias de Benasque Pedro Pascual, IPMU at the University of  Tokyo and INI at Cambridge University for hospitality during the course of this work. This work is supported by the Natural Sciences and Engineering Research Council of Canada.

\end{document}